\documentclass{llncs}
\usepackage{amsmath, amssymb}
\usepackage{amsfonts}
\usepackage{stmaryrd}
\usepackage{alltt}

\title{Automatic Unbounded Verification of Alloy Specifications with Prover9}
\author{Alcino Cunha \and Nuno Macedo}
\institute{{HASLab} --- High Assurance Software Laboratory\\ INESC TEC \& Universidade do Minho, Braga, Portugal\\ \vspace{0.3cm} May 2011}

\begin{document}

\maketitle

\begin{abstract}
  Alloy is an increasingly popular lightweight specification language
  based on relational logic. Alloy models can be automatically
  verified within a bounded scope using off-the-shelf SAT
  solvers. Since false assertions can usually be disproved using small
  counter-examples, this approach suffices for most
  applications. Unfortunately, it can sometimes lead to a false sense
  of security, and in critical applications a more traditional
  unbounded proof may be required. The automatic theorem prover
  Prover9 has been shown to be particularly effective for proving
  theorems of relation algebras~\cite{crp9}, a quantifier-free (or
  point-free) axiomatization of a fragment of relational logic.  In this
  paper we propose a translation from Alloy specifications to fork
  algebras (an extension of relation algebras with the same expressive
  power as relational logic) which enables their unbounded
  verification in Prover9. This translation covers not only logic
  assertions, but also the structural aspects (namely type
  declarations), and was successfully implemented and applied to
  several examples.
\end{abstract}

\section{Introduction}

The Alloy specification language~\cite{alloyb} was created by
following a different approach than the so called ``classic'' formal
methods. It was built to be \emph{lightweight}~\cite{lightweight} and,
instead of focusing on theorem proving, the emphasis is on automatic
analysis. Alloy's underlying logic is a kind of relational logic,
making it easier to write and read specifications without the need to
learn complicated concepts. On the other hand, it is also influenced
by object modeling languages, from which it inherits the navigational
style and type hierarchy. As for the verification of the model, the
Alloy Analyzer tool is provided, which automatically verifies the
specifications within a bounded scope using off-the-shelf SAT solvers.

However, sometimes bounded verification is not enough. In
safety-critical systems, for instance, there is a need to make sure
that the program is \emph{always} correct, i.e., we must perform
\emph{unbounded} verification. The only way to do this is to
mathematically prove that the specifications are correct. Since
Alloy's basic elements are relations, relational logic provides a
natural framework to reason about their specifications. Moreover, we
believe that using a \emph{point-free} (PF) notation, a style where
there are no variables or quantifications, as opposed to
\emph{point-wise} (PW), provides a simpler framework where proofs can
be carried out by simple equational steps (see
e.g.~\cite{escpf}). Fork algebras are the point-free counterpart to
relational logic. They extend the more traditional relation algebras
with products in order to regain the expressiveness of relational
logic, and will be our logic of choice to reason about Alloy
specifications.

Being as expressive as first-order logic, fork algebras are also
\emph{undecidable}: in principle this would restrict us to perform
manual proofs, eventually assisted by interactive theorem
provers. However, off-the-shelf automatic theorem provers (ATPs) are
becoming increasingly efficient and are nowadays able to solve complex
problems. Prover9~\cite{prover9} is an ATP for first-order logic and
equational logic. Studies have show that it is the most efficient
off-the-shelf ATP to deal with relation algebras~\cite{atmcomp}, and
it has been used to prove several properties about them~\cite{crp9}.
Building on these results, we propose a framework for automatic
unbounded verification of Alloy specifications using Prover9. The key
contribution of this framework is a translation from Alloy models into
fork algebras, that covers not only logic assertions, but also other
structural aspects of the model, such as type declarations.

The next section briefly presents Alloy with an
example. Sect.~\ref{sec:fork} presents fork algebras (assuming
previous knowledge of relation algebras) and describes how relations
with arbitrary arity can be represented in this
formalism. Sect.~\ref{sec:alloyforms} and~\ref{sec:alloydecls}
present the translation from Alloy models to fork algebras: the former
focuses on formulas and the latter on the structural
aspects. Sect.~\ref{sec:example} describes the implementation of the
translation, and presents the result of translating our
running example. Sect.~\ref{sec:related} discusses some related
work, and Sect.~\ref{sec:conc} concludes the paper with some
reflexions and suggestions for future work.

\section{Alloy}
\label{sec:alloy}

We will briefly present the Alloy language by following a very simple
example, a model of a university with students and the courses they have completed, which is presented in
Fig.~\ref{fig:ex}. Roughly, an Alloy model is divided in two
parts. The first, the \emph{signature} declarations, defines the
existing types and the relations between them. In our example,
\texttt{Student} and \texttt{Professor} both extend the signature
\texttt{Person}, inducing a type hierarchy. \texttt{Person} is also
declared as abstract, meaning that there are no persons besides those
contained in its sub-signatures.  A particularity of Alloy relations
is that they can have arbitrary arity. For instance, \texttt{course}
is a ternary relation that, for a particular university, relates
students with the courses they have completed. We can also attach
\emph{multiplicities} to signature and relation declarations. For
example, the multiplicity \texttt{some} in relation \texttt{lecturer}
forces that at least one professor is lecturing each course. The
second part of the model consists of \emph{facts}, that define
constraints and properties of the model, and the \emph{assertions} we
want to verify.  
\begin{figure}[t]
\centering
\begin{verbatim}
abstract sig Person {}
sig Student, Professor extends Person {}
sig Course {
  lecturer : some Professor,
  depends : set Course
}
sig University {
  enrolled : set Student,
  courses  : Student -> Course
}
pred inv[u : University] {
  (u.courses).Course in u.enrolled
  all s : Student | (s.(u.courses)).*depends in s.(u.courses)
}
pred enroll[u, u' : University, s : Student] {
  u'.enrolled = u.enrolled + s
  u'.courses = u.courses
}
assert {
  all u,u':University,s:Student | inv[u] and enroll[u,u',s] => inv[u']
}
\end{verbatim}
\caption{Alloy example}
\label{fig:ex}
\end{figure}

\emph{Predicates} can be defined to be reused in facts and
assertions. In our example, we define one predicate to model the
invariant of the university (students with completed courses must be
enrolled in the university, and to have completed a course a student must also complete the courses it depends on), and
another to model an operation which enrolls a new student to the
university. The assertion to be verified in this case is that the invariant is
preserved by the operation.

In order to make translation of the Alloy models easier, we restrict
the grammar to an essential core (see Appendix~\ref{sec:alloysyntax}), to
which almost every other constructions can trivially be reduced.

\section{Fork algebras}
\label{sec:fork}
\emph{Relational logic} (RL) is a characterization of first-order logic (FOL) with relational operators. Although it introduces different notations, RL is as expressive as FOL. If we remove all quantifiers (and variables) from RL, we obtain the \emph{calculus of relations} (CR), a PF fragment of RL. In CR, formulas consist of boolean combinations of inequations, formulas of the form $R \subseteq S$. This calculus is axiomatized by \emph{relation algebras} (RAs) (for a standard axiomatization see~\cite{maddux}), which however is not as expressive as RL (only to a fragment with 3 variables).

\emph{Fork algebras} (FAs)~\cite{forkb} were created to overcome this
expressiveness limitation. They extend CR by introducing pairing
(products) to the base set of the algebra, and a new operator
\emph{fork}, denoted by $\nabla$, and defined by $(x,y) \, (R \nabla S) \, z \equiv x \, R \, z \wedge y \, S \, z$.
FAs are as expressive as RL and their axiomatization, an extension of RA, is the following.

\begin{definition}[Closure Fork Algebras]
A \emph{fork algebra} is an algebraic structure 
\begin{equation*}
\langle U, \cup, \cap, {}^{-}, \bot, \top, \cdot , id, {}^\circ, \nabla \rangle
\end{equation*}
where $\langle U, \cup, \cap, {}^{-}, \bot, \top, \cdot, id, {}^\circ \rangle$ is a relation algebra and for all $R, S, T, Q \in U$,
\begin{align*}
&R \nabla S = ((id \nabla \top) \cdot R) \cap ((\top \nabla id) \cdot S) \\ 
&(R \nabla S)^\circ \cdot (T \nabla Q) = (R^\circ \cdot T) \cap (S^\circ \cdot Q) \\ 
&(id \nabla \top)^\circ \nabla (\top \nabla id)^\circ \subseteq id \\ 
& R^* = id \cup R^* \cdot R \\
& \top \cdot S \cdot R^* \subseteq \top \cdot S \cup (\overline{\top \cdot S} \cap \top \cdot S \cdot R) \cdot R^*
\end{align*}
\label{def:afa}
\end{definition}
The relations $(id \nabla \top)^\circ$ and $(\top \nabla id)^\circ$ select
the first and second element from a pair, and will be denoted by
$\pi_1$ and $\pi_2$, respectively. An operator $\times$ that applies
two relations in parallel can be defined as $R \times S \triangleq (\pi_1 \cdot R) \nabla (\pi_2 \cdot S)$.

\subsection{Handling relations of arbitrary arity}
\label{sec:nary}

While in Alloy relations can have an arbitrary arity, FA is restricted to binary relations. As such, a mechanism must be devised to ``binarize'' arbitrary relations. Unary relations (sets) can be represented in FA using coreflexives, i.e., fragments of the identity relation that filter elements of the given set. More precisely, a unary relation $\mathtt{R}$ can represented by a coreflexive $\Phi_R \subseteq id$ such that $x \in \mathtt{R}$ iff $x\ \Phi_R\ x$. For $n$-ary relations with $n > 2$ a mechanism analogous to uncurrying can be used: an Alloy relation $\mathtt{R} : A_1 \rightarrow \dots \rightarrow A_n$ can be represented by the binary relation $R : A_n \times \dots \times A_2 \rightarrow A_1$, whose domain is the nested product (associated to the right) of the last $n-1$ columns of $\mathtt{R}$. Domains and ranges appear reversed in the binary version to allow a direct encoding of composition: binary relations $\mathtt{R} : A \rightarrow B$ and $\mathtt{S} : B \rightarrow C$ can be composed in Alloy as $\mathtt{R \,.\, S} : A \rightarrow C$ using the dot join; by reversing domains and ranges, this expression can be directly translated to the FA composition $R \cdot S$. Throughout the presentation, the binary representation of an $n$-ary Alloy relation $\mathtt{R}$ will be denoted as $\Phi_R$ if $n=1$ or just $R$ if $n>1$.  We will often abuse the notion of arity and classify the binary version of an $n$-ary relation also as $n$-ary. Given a relation $\mathtt{R}$ its arity will be denoted by $|R|$.

Notice that in Alloy composition is not limited to binary relations. For example, relations $\mathtt{R}: A_1 \rightarrow \dots \rightarrow A_n$ and $\mathtt{S}: B_1 \rightarrow \dots \rightarrow B_m$ can be composed as $\mathtt{R \cdot S} : A_1 \rightarrow \dots \rightarrow A_{n-1} \rightarrow B_2 \rightarrow \dots \rightarrow B_m$, provided $n+m > 2$. In order to translate this directly it is convenient to have an $n$-ary composition operator in FA. Given two relations $R$ and $S$, where $R$ denotes a $n$-ary relation with $n>1$, the composition of $R$ after $S$ will be denoted by $R \bullet^n S$, and defined as follows:
\begin{equation*}
R \bullet^n S = \left\{ \begin {array}{cl}
R \cdot S & \textrm{if $n = 2$}\\
R \bullet^{n-1} (id \times S) & \textrm{if $n > 2$}
\end {array} \right.
\label{eq:uncomp}
\end{equation*}
It is trivial to show by induction that $n$-ary composition satisfies analogous properties to normal binary composition, in particular identity and associativity: 
\begin{align*}
 & R \bullet^n id = R \quad \wedge \quad id \bullet^2 R = R\\
 & R \bullet^n (S \bullet^m T) = (R \bullet^n S) \bullet^{n+m-2} T
\end{align*}
These properties enable us to calculate directly with this operator without the the need to expand its definition. Composition with unary relations can be performed with normal binary composition since these are represented as binary coreflexives.

\newcommand{\rotate}[1]{\overrightarrow{#1}}
\newcommand{\nrotate}[2]{\overset{#2 \rightarrow}{#1}}
In an $n$-ary relation the notion of range and domain is somehow arbitrary: for example, given a ternary relation $\mathtt{R} : A \rightarrow B \rightarrow C$ we may want to compose it via the middle column with a relation $\mathtt{S} : B \rightarrow D$. To allow this, we will define an operator to rotate a relation: given an $n$-ary relation $R : A_n \times \dots \times A_2 \rightarrow A_1$, its right-rotation will be denoted by $\rotate{R} : A_{n-1} \times \dots \times A_1 \rightarrow A_n$ and can be defined as
\begin{equation*}
\rotate{R} = X_{n-1}^{n -1} \cdot (R \nabla  X_1^{n - 1} \nabla \dots \nabla  X^{n - 1}_{n-2})^\circ 
\end{equation*}
where $(R \nabla \dots \nabla  S)$ represents the right-nested fork
$R \nabla (\dots \nabla S))$, and $X^n_i$ selects the $i$th component of a right-nested $n$-ary tuple, according to the following definition:
\begin{equation*} 
X^n_i = \left\{ \begin{array}{ll} 
id & \textrm{if } n = i = 1\\ 
\pi_1 & \textrm{if } n > 1 \wedge i = 1 \\
X^{n-1}_{i-1} \cdot \pi_2 & \textrm{if } n > 1 \wedge i > i 
\end{array} \right. 
\label{eq:defxi}
\end{equation*} 
Note that, when $R$ is a binary relation, $\rotate{R} = R^\circ$. We will denote by $\nrotate{R}{k}$ the application of the rotate $k$ times. Likewise to $n$-ary composition, this operator satisfies some useful calculational properties, namely:
\begin{equation*}
\nrotate{R}{|R|} = R \quad \wedge \quad \nrotate{R \bullet^{|R|} S}{(|R|-1)} \equiv \rotate{S} \bullet^{|S|} \nrotate{R}{(|R|-1)}
\end{equation*}

\section{Translating Alloy formulas to FA}
\label{sec:alloyforms}

While in RL variables can only be used in relation application, i.e
membership testing, in Alloy they can be used as normal unary
relations in relational formulas. For example, we can have the
composition $\mathtt{R . x . S}$, where \texttt{x} denotes a variable,
and \texttt{R} and \texttt{S} arbitrary relational expressions. This
feature makes the direct translation of Alloy formulas to FA quite
difficult, since standard heuristics for variable elimination cannot
be used. 
To overcome this problem, our approach is to first expand Alloy
formulas to standard RL, where variables only appear applied to
relations, and then use a RL to FA translation to remove the
variables. As will be seen in Sect.~\ref{sec:heuristics}, this allows
us to use powerful heuristics for variable elimination and output much
simpler FA formulas.

\subsection{From Alloy formulas to RL}
\label{sec:tralloy}

Fig.~\ref{fig:alloy2rl} presents the translation of Alloy formulas to
RL. Alloy formulas are boolean combinations of atomic formulas of
shape $\mathtt{some\ X}$ or $\mathtt{x\ \texttt{in}\ X}$, where
$\mathtt{X}$ and $\mathtt{Y}$ denote Alloy relational formulas. The
first set of rules in Fig.~\ref{fig:alloy2rl} introduces variables in
order to convert these atomic formulas into relation applications of
shape $x \in \mathtt{X}$, where $x$ denotes a tuple of variables and
$\mathtt{X}$ is still an Alloy relational formula. The second set of
rules expands these into boolean combinations of standard relation
applications by computing the expected semantics of relational
operators. Moreover, variable occurrences are translated to equality
tests (using the identity relation), and constant unary relations
(denoting signatures) to the corresponding coreflexives that filter
values of that type.

\begin{figure}[t]
  \centering
\begin{align*}
&\llbracket \mathtt{! R} \rrbracket  \equiv \neg \llbracket \mathtt{R} \rrbracket  \\
&\llbracket \mathtt{R\ \&\&\ S} \rrbracket \equiv \llbracket \mathtt{R} \rrbracket \wedge \llbracket \mathtt{S} \rrbracket  \\
&\llbracket \mathtt{all\ x : X \, \texttt{|} \, S} \rrbracket \equiv \langle \forall x : \llbracket \mathtt{x\ \texttt{in}\ X} \rrbracket : \llbracket \mathtt{S} \rrbracket \rangle \\
&\llbracket \mathtt{some\ X} \rrbracket \equiv \langle \exists x_1,\dots,x_{|\mathtt{X}|} :: \llbracket (x_1,\dots,x_{|\mathtt{X}|}) \in \mathtt{X} \rrbracket \rangle   \\
&\llbracket \mathtt{X\ \texttt{in}\ Y} \rrbracket \equiv \langle \forall x_1,\dots,x_{|\mathtt{X}|} : \llbracket (x_1,\dots,x_{|\mathtt{X}|}) \in \mathtt{X} \rrbracket : \llbracket (x_1,\dots,x_{|\mathtt{X}|}) \in \mathtt{Y} \rrbracket \rangle  \\
&\\
&\llbracket x \in \mathtt{v} \rrbracket \equiv x\, id\, v  \\
&\llbracket (x_1,\dots,x_{|\mathtt{X}|}) \in \mathtt{R} \rrbracket \equiv x_1 \, R \, (x_{|\mathtt{2}|},\dots,x_{|\mathtt{X}|})  \\
&\llbracket x \in \mathtt{S} \rrbracket \equiv x \, \Phi_S \, x  \\
&\llbracket x \in \mathtt{univ} \rrbracket \equiv \mathit{true}   \\
&\llbracket x \in \mathtt{none} \rrbracket \equiv \mathit{false}   \\
&\llbracket (x_1,x_2) \in \mathtt{iden} \rrbracket \equiv x_1\, id\, x_2   \\
&\llbracket (x_1,\dots,x_{|\mathtt{X}|}) \in \mathtt{X\ \texttt{+}\ Y} \rrbracket \equiv \llbracket (x_1,\dots,x_{|\mathtt{X}|}) \in \mathtt{X} \rrbracket \vee \llbracket (x_1,\dots,x_{|\mathtt{X}|}) \in \mathtt{Y} \rrbracket \\
&\llbracket (x_1,\dots,x_{|\mathtt{X}|}) \in \mathtt{X\ \texttt{\&}\ Y} \rrbracket \equiv \llbracket (x_1,\dots,x_{|\mathtt{X}|}) \in \mathtt{X} \rrbracket \wedge \llbracket (x_1,\dots,x_{|\mathtt{X}|}) \in \mathtt{Y} \rrbracket  \\
&\llbracket (x_1,\dots,x_{|\mathtt{X}|}) \in \mathtt{X\ \texttt{-}\ Y} \rrbracket \equiv \llbracket (x_1,\dots,x_{|\mathtt{X}|}) \in \mathtt{X} \rrbracket \wedge \neg \llbracket (x_1,\dots,x_{|\mathtt{X}|}) \in \mathtt{Y} \rrbracket  \\
&\llbracket (x_1,x_2) \in \mathtt{\texttt{\textasciitilde}\ X} \rrbracket \equiv \llbracket (x_2,x_1) \in \mathtt{X} \rrbracket   \\
&\llbracket (x_1,\dots,x_{|\mathtt{X}|+|\mathtt{Y}|-2}) \in \mathtt{X\ \texttt{.}\ Y} \rrbracket \equiv \langle \exists k :: \llbracket (x_1,\dots,x_{|\mathtt{X}|-1},k) \in \mathtt{X} \rrbracket \wedge \llbracket (k,x_{|\mathtt{X}|},\dots,x_{|\mathtt{X}|+|\mathtt{Y}|-2}) \in \mathtt{Y} \rrbracket \rangle  \\
&\llbracket (x_1,\dots,x_{|\mathtt{X}|+|\mathtt{Y}|}) \in \mathtt{X\ \texttt{->}\ Y} \rrbracket \equiv   \llbracket (x_1,\dots,x_{|\mathtt{X}|}) \in \mathtt{X} \rrbracket \wedge \llbracket (x_{|\mathtt{X}| + 1},\dots,x_{|\mathtt{X}|+|\mathtt{Y}|})  \in \mathtt{Y} \rrbracket  \\ 
&\llbracket (x_1,\dots,x_{|\mathtt{Y}|}) \in \mathtt{X\ \texttt{<:}\ Y} \rrbracket \equiv \llbracket x_1 \in \mathtt{X} \rrbracket \wedge \llbracket (x_1,\dots,x_{|\mathtt{Y}|}) \in \mathtt{Y} \rrbracket  \\
&\llbracket (x_1,\dots,x_{|\mathtt{X}|}) \in \mathtt{X\ \texttt{:>}\ Y} \rrbracket \equiv \llbracket x_{|\mathtt{X}|} \in \mathtt{Y} \rrbracket \wedge \llbracket (x_1,\dots,x_{|\mathtt{X}|}) \in \mathtt{X} \rrbracket  
\end{align*}  
  \caption{Translation from Alloy formulas to RL}
  \label{fig:alloy2rl}
\end{figure}

\subsection{From RL to FA}
\label{sec:trpwpf}

The translation from RL to FA will be defined as a \emph{strategic
  rewriting} process~\cite{sw}: basic rewrite rules are combined using
strategic combinators in order to build powerful rewrite systems. We
will use the following combinators: \emph{sequence}, a binary
combinator denoted by $\triangleright$, which applies the second rule
to the result of the first, if successful; \emph{choice}, a binary
combinator denoted by $\oslash$, which applies the first rule or, if it
fails, applies the second; \emph{many}, which applies a rule
repetitively until it fails; and \emph{once}, which applies a rule once
somewhere inside a term.

For example, we can easily define a strategy to eliminate universal
quantifiers and implications as follows:
\begin{align}
\phi \Rightarrow \psi & \rightsquigarrow \neg \phi \vee \psi \label{eq:impldef} \\
\langle \forall z :: \phi \rangle & \rightsquigarrow \neg \langle \exists z :: \neg \phi \rangle \label{eq:trdef2}\\
\langle \forall z : \phi : \psi \rangle & \rightsquigarrow \langle \forall z :: \phi \Rightarrow \psi \rangle \label{eq:rangeelim}
\end{align}
\begin{equation*}
\mathit{normalize} \triangleq \mathit{many} \, (\mathit{once} \, \eqref{eq:impldef} \oslash \mathit{once} \, \eqref{eq:trdef2} \oslash \mathit{once} \, \eqref{eq:rangeelim})
\end{equation*}
To distinguish arbitrary RL formulas from purely relational
expressions, variables $\phi$, $\psi$, \ldots, will be used to denote
the former, and variables $R$, $S$, \ldots, to denote the latter.

Intuitively, the translation of a normalized RL formula to an
equivalent FA formula will proceed as follows:
\begin{enumerate}
\item The formula will first be universally quantified over a fresh
  pair of special variables, denoted $\mathbf{x}$ and $\mathbf{y}$, that will represent
  arbitrary output and input values, respectively. The following rule
  performs this transformation.
  \begin{equation*}
    \mathit{insertVars} \triangleq \phi \rightsquigarrow \langle \forall \mathbf{x},\mathbf{y} :: \phi \rangle
  \end{equation*}
\item Together with quantifier elimination, all relation applications
  will iteratively be uniformed to operate on the same variables, so
  that boolean combinations of applications can be combined into a
  single application using the following simple strategy:
  \begin{align}
    u \, R \, v \wedge u \, S \, v & \rightsquigarrow u \, (R \cap S) \, v \label{eq:aux1} \\
    u \, R \, v \vee u \, S \, v & \rightsquigarrow u \, (R \cup S) \, v \label{eq:aux2} \\
    \neg (u \, R \, v) & \rightsquigarrow u \, \overline{R} \, v \label{eq:aux3}
  \end{align}
  \begin{equation*}
    \mathit{aggregate} \triangleq \mathit{once} \, \eqref{eq:aux1} \oslash \mathit{once} \, \eqref{eq:aux2} \oslash \mathit{once} \, \eqref{eq:aux3}
  \end{equation*}
  Here, $u$ and $v$ denote arbitrary variables or variable tuples.
\item In the end we will obtain a single relational expression applied
  to the quantified special input and output variables. Those can then
  be eliminated by the following rule:
  \begin{equation*}
    \mathit{dropVars} \triangleq \langle \forall u,v :: u \, R \, v \rangle \rightsquigarrow R = \top
  \end{equation*}

\end{enumerate}
We will now detail the strategies for uniforming applications and
existential quantifier elimination.

\paragraph{Uniforming Applications.} To simplify the presentation we
will represent variables by indices (similar to de Bruijn indices):
existential quantifiers do not explicitly mention variable names and a
quantified variable is referred to by the natural number that indexes the
respective quantifier. For example, the formula $\langle \exists ::
\langle \exists :: 2\, R\, 1 \rangle \rangle$ is equivalent to
$\langle \exists x_1 :: \langle \exists x_2 :: x_2 \, R \, x_1 \rangle
\rangle$. The depth of an application is the number of existential
quantifiers that enclose it. The depth of a formula is the maximum
depth of all applications. In the following presentation, $n$ denotes
the depth of the application that matches with the left hand side of
a rewrite rule.

To uniform the input of applications we generalize it to a tuple
containing all existentially quantified variables at that depth, and
use the selection operator to choose the desired variable:
\begin{equation*}
i \, R \, j = i \, (R \cdot X^n_j) \, (1, \dots, n)
\end{equation*}

In order to keep the tuple of quantified variables only on the input
side, a slightly different strategy is used to uniform the output,
motivated by the following calculation:
\begin{align*}
  i\, (R \cdot X^n_j) \, (1, \dots, n) & = \langle \exists z :: z \, X^n_i \, (1, \dots, n) \wedge z \, (R \cdot X^n_j) \, (1, \dots, n) \rangle\\
  & = \langle \exists z :: z \, (X^n_i \cap (R \cdot X^n_j)) \, (1, \dots, n) \rangle\\
  & = \langle \exists z :: \mathbf{x} \, \top \, z \wedge z \, (X^n_i \cap (R \cdot X^n_j)) \, (1, \dots, n) \rangle\\
  & = \mathbf{x} \, (\top \cdot (X^n_i \cap (R \cdot X^n_j))) \, (1, \dots, n)
\end{align*}
The quantified (temporary) variable $z$ is introduced to bind the
output with the appropriate variable of the input tuple. The
quantification is then eliminated by introducing an intersection,
composing with $\top$, and applying the resulting expression to the
special variable $\mathbf{x}$. A strategy to uniform applications at
any given depth can thus be implemented as follows:
\begin{equation*}
\mathit{uniform} \triangleq \mathit{once} \, (i \, R \, j \rightsquigarrow \mathbf{x} \, (\top \cdot (X^n_i \cap (R \cdot X^n_j))) \, (1, \dots, n))
\end{equation*}
Although this strategy only covers applications to a single variable,
it is rather straightforward to generalize it to applications to tuples of
variables using forks of projections, as described in~\cite{allpf}. 

\paragraph{Removing Existential Quantifiers.} The strategy to remove
existential quantifiers is similar to the one used above to remove the
temporary quantifier. While $n > 1$, we compose the terms with the
relation $\langle id, \top \rangle$ on the right side to cut the
right-most existential quantified variable from the input variable tuple:
\begin{equation}
\langle \exists :: \mathbf{x} \, R \, (1, \dots, n) \rangle \rightsquigarrow \mathbf{x} \, (R \cdot \langle id, \top \rangle) \, (1, \dots, n-1) \label{eq:trdef91}
\end{equation}
If $n = 1$, the term is composed with $\top$ and applied to the special variable $\mathbf{y}$:
\begin{equation}
\langle \exists :: \mathbf{x} \, R \, 1 \rangle \rightsquigarrow \mathbf{x} \, (R \cdot \top) \, \mathbf{y} \label{eq:trdef92}
\end{equation}
\begin{equation*}
\mathit{dropExists} \triangleq \mathit{once} \, \eqref{eq:trdef91} \oslash \mathit{once} \, \eqref{eq:trdef92}
\end{equation*}

Uniforming applications enables the application of strategy
$aggregate$, reducing all applications at the deepest level of the
formula to a single application. This in turn allows strategy
$dropExists$ to remove one level of quantifiers, thus reducing the
depth of the formula. These steps can be combined in the following strategy:
\begin{equation*}
  \mathit{shorten} \triangleq \mathit{uniform} \triangleright \mathit{aggregate} \triangleright \mathit{dropExists}
\end{equation*}

Finally, the translation from RL to FA iteratively shortens the depth
of the formula until the external universally quantified variables can
be dropped.
\begin{align*}
\mathit{translate} \triangleq \mathit{normalize} \triangleright \mathit{insertVars} \triangleright \mathit{many} \, (\mathit{dropVars} \oslash \mathit{shorten})
\end{align*}

\subsection{Heuristic simplification}
\label{sec:heuristics}

An heuristic rewrite strategy can be defined to further simplify the
resulting FA expressions. This simplification is based on two key
ideas: the relaxation of the form of the resulting formula, which we
now allow to be an inequation of the form $R \subseteq S$; and the use of
additional relational operators in order to remove quantifiers. We
will only present a simplified version of this strategy: the full set
of rules included in the implementation can be found in~\cite{allpf}.

The heuristic simplification strategy is defined as
\begin{equation*}
\mathit{simplify} \triangleq \mathit{many} \, (\mathit{FOLrules} \oslash \mathit{definitions} \oslash \mathit{FArules})
\end{equation*}
where
\begin{itemize}
\item $\mathit{FOLrules}$ performs several simplifications related to
  first-order logic. Essentially, it is the \emph{choice} of rules like those
  presented in Table~\ref{tab:logicrules}, applied \emph{once}. Note
  that we present simplified versions of the rules. The implemented
  version considers the commutativity and associativity of logical
  operators. Since we allow arbitrary inequations as result, this
  strategy also tries to push some expressions into the range of
  universal quantifiers, thus distributing the complexity of the
  formula between both sides of the inequation.
\item $\mathit{definitions}$ applies definitions of relational
  operators in order to remove quantifiers. For example, it applies
  the definition of composition and converse in order to remove some
  existential quantifiers, according to the following rules:
  \begin{align*}
    \langle \exists w :: u \, R \, w \wedge w \, S \, v \rangle & \rightsquigarrow u \, (R \cdot S) \, v \\
    \langle \exists w :: u \, R \, w \wedge v \, S \, w \rangle & \rightsquigarrow u \, (R \cdot S^\circ) \, v \\
    \langle \exists w :: w \, R \, u \wedge w \, S \, v \rangle & \rightsquigarrow u \, (R^\circ \cdot S) \, v
  \end{align*}
  In particular, it introduces the $n$-ary composition operator
  introduced in Sect.~\ref{sec:nary} whenever possible, by
  generalizing the rules above. Among others, we have the following
  rule:
  \begin{align*}
    \langle \exists x_1 :: u \, R \, (x_1,\dots,x_n) \wedge x_n \, S \, v \rangle & \rightsquigarrow u \, (R \bullet^n S) \, (x_1,\dots,x_{n-1},v)
  \end{align*}
  Expressions similar to this one, but with $x_1$ on a different position of the tuple are dealt with other rules, which resort to the rotate operator.
  Besides composition, we also introduce forks, complement, and other
  derived operators characterized by powerful algebraic laws, such as
  the relational division operators. These are particularly
  interesting because they allow us to remove universal quantifiers,
  according to the following rules:
  \begin{align*}
   \langle \forall w : w \, R \, u : w \, S \, v \rangle & \rightsquigarrow u \, R \backslash S \, v\\
   \langle \forall w : u \, R \, w : v \, S \, w \rangle & \rightsquigarrow u \, R / S \, v
  \end{align*}
\item $\mathit{FArules}$ performs several simplifications related to
  FA operators. Among many others, it is the \emph{choice} of all
  rules presented in Table~\ref{tab:rlrules}, applied
  \emph{once}. These rules correspond to equations of the
  axiomatization of FA presented in Sect.~\ref{sec:fork} or can
  easily be derived from them.
\end{itemize}

\begin{table}[t]
\center
\begin{tabular}{|c|c|c|}
\hline
Conjunction & Disjunction & Implication  \\
\hline
$\psi \wedge \mathit{true} \rightsquigarrow \psi$ & $\psi \vee \mathit{false} \rightsquigarrow \psi$ & $\neg \psi \vee \phi \rightsquigarrow \psi \Rightarrow \phi$\\
$\psi \wedge \mathit{false} \rightsquigarrow \mathit{false}$ & $\psi \vee \mathit{true} \rightsquigarrow \mathit{true}$ &  $\mathit{false} \Rightarrow \psi \rightsquigarrow \mathit{true}$\\
$\psi \wedge \psi \rightsquigarrow \psi$ & $\psi \vee \psi \rightsquigarrow \psi$ & $\mathit{true} \Rightarrow \psi \rightsquigarrow \psi$\\
$\neg (\psi \wedge \phi) \rightsquigarrow \neg \psi \vee \neg \phi$ & $ \neg (\psi \vee \phi) \rightsquigarrow \neg \psi \wedge \neg \phi$ & $\psi \Rightarrow (\phi \Rightarrow \varphi) \rightsquigarrow (\psi \wedge \phi) \Rightarrow \varphi$ \\
\hline
Negation &\multicolumn{2}{|c|}{Quantifiers} \\
\hline
$\neg \neg \psi \rightsquigarrow \psi$ & \multicolumn{2}{|c|}{$\langle \forall u : \psi : \phi \Rightarrow \varphi \rangle \rightsquigarrow \langle \forall u : \psi \wedge \phi : \varphi \rangle$}\\
$\neg \mathit{true} \rightsquigarrow \mathit{false}$ & \multicolumn{2}{|c|}{$\langle \exists u : \psi : \phi  \rangle \rightsquigarrow \langle \exists u :: \psi \wedge \phi \rangle$} \\
$\neg \mathit{false} \rightsquigarrow \mathit{true}$ & \multicolumn{2}{|c|}{$\langle \forall u : \psi : \langle \forall v : \phi : \varphi \rangle \rangle \rightsquigarrow \langle \forall u,v : \psi \wedge \phi : \varphi \rangle$} \\
& \multicolumn{2}{|c|}{$\langle \exists u : \psi : \langle \exists v : \phi : \varphi \rangle \rangle \rightsquigarrow \langle \exists u,v : \psi \wedge \phi : \varphi \rangle$} \\
\hline
\end{tabular}
\caption{FOL rules.}
\label{tab:logicrules}
\end{table}

\begin{table}[t]
\center
\begin{tabular}{|c|c|c|}
\hline
Meet rules & Join rules & Complement rules\\
\hline
$A \cap A \rightsquigarrow A$ & $A \cup A \rightsquigarrow A$ & $\overline{R \cdot S} \rightsquigarrow R^\circ \backslash \overline{S}$  \\
$A \cap \top \rightsquigarrow A$ & $A \cup \top \rightsquigarrow \top$ & $\overline{R \backslash S} \rightsquigarrow R^\circ \cdot \overline{S}$ \\
$A \cap \bot \rightsquigarrow \bot$ & $A \cup \bot \rightsquigarrow A$ & $\overline{S} \subseteq \overline{R} \rightsquigarrow R \subseteq S$ \\
$(A \cap B)^\circ \rightsquigarrow A^\circ \cap B^\circ$ & $(A \cup B)^\circ \rightsquigarrow A^\circ \cup B^\circ$ &  $\overline{\overline{R}} \rightsquigarrow R$  \\
$\overline{A \cup B} \rightsquigarrow \overline{A} \cap \overline{B}$ & $\overline{A \cap B} \rightsquigarrow \overline{A} \cup \overline{B}$ &  $\overline{\overline{R}^\circ} \rightsquigarrow R^\circ$ \\
\hline
Converse rules & Division rules & Product rules\\
\hline
$R^{\circ^\circ} \rightsquigarrow R$ & $R \subseteq S \backslash T  \rightsquigarrow S \cdot R \subseteq T$ & $\pi_1^\circ \cdot R \cap \pi_2^\circ \cdot S  \rightsquigarrow (R \nabla S) \rangle$ \\
$(R \cdot S)^\circ \rightsquigarrow S^\circ \cdot R^\circ$ & $\bot \backslash R \rightsquigarrow \mathit{true}$ & $(R \nabla S)^\circ \cdot (A \nabla B) \rightsquigarrow R^\circ \cdot A \cap S^\circ \cdot B $\\
$\top \subseteq R^\circ \rightsquigarrow \top \subseteq R$ & $R \backslash \top \rightsquigarrow \mathit{true}$ &  $(id \nabla \top) \cdot R \rightsquigarrow (R  \nabla \top \cdot R)$ \\
$R^\circ \subseteq \bot  \rightsquigarrow R \subseteq \bot$ & $id \backslash R \rightsquigarrow R$ & $(R \cdot \pi_1 \nabla  S \cdot \pi_2) \rightsquigarrow R \times S$ \\
\hline
\end{tabular}
\caption{FA rules.}
\label{tab:rlrules}
\end{table}

With this strategy we can redefine the translation from Alloy to FA as follows:
\begin{align*}
\mathit{translate} \triangleq \mathit{simplify} \triangleright (\mathit{dropVars} \oslash (&\mathit{normalize} \triangleright \mathit{insertVars} \, \triangleright \\ & \mathit{many} \, (\mathit{dropVars} \oslash (\mathit{simplify} \triangleright \mathit{shorten}))))
\end{align*}
By applying the heuristic strategy we highly simplify the formula
before the automatic translation kicks in. At this point, the
expression might even be in a shape where variables can be
dropped. Since it keeps the range in universal quantifiers, the
$\mathit{dropVars}$ strategy must be redefined as follows:
\begin{equation*}
  \mathit{dropVars} \triangleq (\langle \forall u,v :: u \, R \, v \rangle \rightsquigarrow R = \top) \oslash (\langle \forall u,v : u \, R \, v : u \, S \, v\rangle \rightsquigarrow R \subseteq S)
\end{equation*}

\subsection{Translating the closures}
Translating the reflexive-transitive and transitive closures requires a slightly different approach. In this section only the translation of the reflexive-transitive closure will be presented, but a similar translation can be applied to the reflexive closure. Due to the closure type restrictions, our goal is to obtain an expression of the form:
\begin{equation}
	\llbracket (x,y) \in \mathtt{*R} \rrbracket \equiv (a_1,\dots,a_k,x) \, (\mathit{translate*} \llbracket (x,y) \in \mathtt{R} \rrbracket)^* \, (a_1,\dots,a_k,y)
\end{equation}
Where $(a_1,\dots,a_k)$ are the $k$ free variables occurring in $\mathtt{R}$, with type $A_1 \times \dots \times A_k$. This operation ``lifts'' the variable applications from inside the closure, resulting in the following type transformation:
\begin{equation*}
	\mathtt{R} \, (a_1,\dots,a_k) : A \leftarrow A \rightsquigarrow \mathit{translate*} \llbracket (x,y) \in \mathtt{R} \rrbracket : A_1 \dots A_k \times A \leftarrow A_1 \dots A_k \times A 
\end{equation*}
The resulting expression can then be processed by the standard PF transformation already defined, where the variables will be dropped.

The expansion of the inside expression by the $\llbracket \, \rrbracket$ operation results only in the introduction of boolean operators and existential quantifiers, which have to be removed in order to obtain the desired PF expression. The existential quantifications may be removed by the \emph{definitions} rule, which applies the composition or $n$-ary composition, associated with the reverse or rotate operators, to drop those variables. In order to obtain the wanted shape of input and output, we apply a rule similar to \emph{uniform}, but with a tuple on each side instead:
\begin{equation*}
\mathit{uniform*} \triangleq \mathit{many} \, (\mathit{once} \, (a_i \, R \, a_j \rightsquigarrow (a_1, \dots, a_n, x) \, ({X^{n}_i}^\circ \cdot R \cdot X^{n}_j) \, (a_1, \dots,a_n, y))) \notag
\end{equation*}
Once again, if $a_i$ or $a_j$ are tuples, this transformation is generalized to forks of $X$ projections. The rest of the boolean operators may then be removed by the \emph{aggregate} rule already defined, by applying intersections, reunions and complements. The translation is thus defined by:
\begin{equation}
	\mathit{translate*} \triangleq \mathit{many} (\mathit{aggregate} \triangleright \mathit{definitions} \triangleright \mathit{uniform*})
\end{equation}

\section{Translating Alloy declarations to FA}
\label{sec:alloydecls}

Besides translating Alloy formulas to FA we also need to express in
this formalism other facts about an Alloy model, namely all the
information concerning signature, relation and multiplicity
declaration.

Alloy signatures denote sets of atoms and thus will be represented by
constant coreflexives. A sub-signature declaration induces an
inclusion between its coreflexive and the super-signature
coreflexive. Sub-signatures must also be pair-wise disjoint. If a
signature is abstract, then its coreflexive is exactly the union of
all the coreflexives of its sub-signatures. Moreover, according to
Alloy semantics, the identity relation is equal to the union of all the
top-level signature coreflexives.

A binary relation of type $R :: A \rightarrow B$ induces the fact $R
\subseteq \Phi_B \cdot \top \cdot \Phi_A$. In case of $n$-ary relations, since we transform the domain into a product, the property can be generalized to $R \subseteq \Phi_{A_1} \cdot \top \cdot \Phi_{A_2} \times \dots \Phi_{A_n}$, for a relation $R :: A_1 \rightarrow \dots \rightarrow A_n$. Notice that type reasoning can be easily performed using this kind of declaration. For instance,
since the inclusion is monotone, for two relations $R \subseteq \Phi_A
\cdot \top \cdot \Phi_B$ and $S \subseteq \Phi_C \cdot \top \cdot
\Phi_D$ we have $R \cdot S \subseteq \Phi_A \cdot \top \cdot \Phi_B
\cdot \Phi_C \cdot \top \cdot \Phi_D$. Since for any coreflexives
$\Phi$ and $\Psi$ we have $\Phi \cdot \Psi = \Phi \cap \Psi$, we see
that the values that are composed are precisely the ones belonging to
$\Phi_B \cap \Phi_C$. In particular, if the signatures are disjoint we
have $R \cdot S \subseteq \bot$ as expected.

Signature and relation multiplicities can be used to introduce many
useful constraints in an Alloy model, without the need for additional
facts. They can be directly expressed as FA properties, thus avoiding
the need to be formulated in RL and processed by the default
transformation. For example, a signature $A$ constrained by
multiplicity \texttt{some} induces the property $\top \subseteq \top
\cdot \Phi_A \cdot \top$ involving its coreflexive. The following
calculation makes evident why this formula expresses the desired
meaning:
\begin{align*}
  \top \subseteq \top \cdot \Phi_A \cdot \top & \equiv \langle \forall x, y :: x\, \top\, y \Rightarrow x \, (\top \cdot \Phi_A \cdot \top)\, y \rangle\\
  & \equiv \langle \forall x, y :: \mathit{true} \Rightarrow \langle \exists z, w :: x \, \top \, z \wedge z \, \Phi_A \, w \wedge w \, \top\, y \rangle \rangle\\
  & \equiv \langle \forall x, y :: \langle \exists z, w :: \mathit{true} \wedge z \, \Phi_A \, w \wedge \mathit{true} \rangle \rangle\\
  & \equiv \langle \exists z, w :: z \, \Phi_A \, w \rangle
\end{align*}
A similar calculation allows us to deduce that a signature $A$
constrained by multiplicity \texttt{lone} can be represented the
property $\Phi_A \cdot \top \cdot \Phi_A \subseteq id$. If the
multiplicity is \texttt{one} we just generate both facts.

Concerning relations, each multiplicity in a declaration will be
treated separately and will induce a different FA property. A
\texttt{lone} in the last column of a relation declaration (for
example, $R : A \rightarrow \mathtt{lone}\, B$) can be represented by
the property $R^\circ \cdot R \subseteq id$, as justified by the
following calculation:
\begin{align*}
  R^\circ \cdot R \subseteq id & \equiv \langle \forall x,y :: x\, (R^\circ \cdot R)\, y \Rightarrow x\, id\, y \rangle \\
  &  \equiv \langle \forall x,y :: \langle \exists z :: x\, R^\circ \, z \wedge z \, R\, y \rangle \Rightarrow x = y \rangle \\
  & \equiv \langle \forall x,y :: \langle \exists z :: z\, R \, x \wedge z \, R\, y \rangle \Rightarrow x = y \rangle
\end{align*}
If the \texttt{lone} multiplicity appears in a column other than the
last, we can just apply the rotate operator in combination with the above property. In general, given a \texttt{lone} in the $i$th column we have the following equivalent property:
\begin{equation*}
\nrotate{R}{(|R|-i)}^\circ \cdot \nrotate{R}{(|R|-i)} \subseteq id
\end{equation*}
A similar reasoning can be applied to the \texttt{some}
multiplicity. If the $i$th field is marked with this multiplicity we
generate the following property:
\begin{equation*}
id \subseteq \nrotate{R}{(|R|-i)} \cdot \nrotate{R}{(|R|-i)}^\circ
\end{equation*}
Likewise to signatures, if the multiplicity is \texttt{one} we generate both facts.

\section{Implementation and example}
\label{sec:example}

The translations defined in Sects.~\ref{sec:alloyforms} and
\ref{sec:alloydecls} were implemented in a tool that transforms Alloy
models into FA inequations ready to be proved by Prover9. The tool is
divided in three parts: first Alloy models are parsed and translated
to RL; then the RL models are transformed to FA; lastly the FA model
is embedded into Prover9.

All the translations were implemented in the functional language
Haskell. Parsing of the Alloy model is performed by resorting to the
parser generator \emph{Happy}~\cite{happy}. The model is then
transformed to fit the restricted grammar defined in
Appendix~\ref{sec:alloysyntax}. Finally, by applying the rules defined in
Fig.~\ref{fig:alloy2rl} and the properties induced by the signatures
defined in Sect.~\ref{sec:alloydecls} we obtain the RL model.

In order to implement the translation from RL to FA, defined in
Sect.~\ref{sec:alloyforms}, a generic RL strategic rewriter was
implemented, allowing an effective way to manipulate formulas by
defining rules and strategies. As such, once the needed RL rules were
defined, we were able to define not only the translation from PW to
PF, with several variants, but also its reverse, from PF to PW. The
facts and the properties induced by the signature declarations are
passed on as the environment of the other translations. Finally,
although this tool was implemented with Alloy in mind, a mode for
directly translating RL formulas to FA is also provided.

The output of the translation is a step-by-step RL to FA translation
along with the final FA model, pretty-printed in \LaTeX. In
Fig.~\ref{fig:exfa} the translation of the example Alloy model in
Fig.~\ref{fig:ex} is presented. For space reasons, in the assertion we
abbreviate signature and relation names to the first letter and denote
by $R_i$ the term $(X_i^{3 \, \circ} \cdot R)$. This FA model can now
be passed to Prover9. To prove theorems in Prover9, it is necessary to
define useful valid axioms and lemmas for the particular
context. In order to efficiently verify Alloy models, useful
relational and FA properties were defined. In particular, the
assertion in our example model was automatically verified by Prover9.

\begin{figure}[t]
  \centering
\begin{align*}
&id = \Phi_{\mathit{Person}} \cup \Phi_{\mathit{Course}} \cup \Phi_{\mathit{University}}\\
&\Phi_{\mathit{Student}} \cup \Phi_{\mathit{Professor}} \subseteq \Phi_{\mathit{Person}} \wedge \Phi_{\mathit{Student}} \cap \Phi_{\mathit{Professor}} = \bot\\
&\mathit{lecturer} \subseteq \Phi_{\mathit{Course}} \cdot \top \cdot \Phi_{\mathit{Professor}}\\
&\mathit{enrolled} \subseteq \Phi_{\mathit{University}} \cdot \top \cdot \Phi_{\mathit{Student}}\\
&\mathit{courses} \subseteq \Phi_{\mathit{University}} \cdot \top \cdot \Phi_{\mathit{Student}} \times \Phi_{\mathit{Course}}\\
&\mathit{depends} \subseteq \Phi_{\mathit{Course}} \cdot \top \cdot \Phi_{\mathit{Course}}\\
&id \subseteq lecturer \cdot \mathit{lecturer}^\circ
\end{align*}  
\begin{displaymath}
  \begin{array}{c}
(\Phi_{\mathit{U}} \times \Phi_{\mathit{U}} \times \Phi_{\mathit{S}})
\cap \mathit{c}_1 / (\mathit{e}_1 \cdot \pi_1) \cap
(c_1 \cdot (\Phi_{\mathit{S}} \times d^{*\circ})) / c_1
\\ \cap \\ \mathit{c}_1 /
\mathit{c}_2 \cap \mathit{c}_2 / \mathit{c}_1 \cap
\mathit{e}_1 / \mathit{e}_2 \cap \mathit{e}_2 \cdot \pi_2 \cdot \pi_2
\cap \mathit{e}_2 / (\mathit{e}_1 \cup \mathit{id}_3)\\
\subseteq\\ 
 \mathit{c}_2 / (\mathit{e}_2 \cdot \pi_1)  \cap
(c_2 \cdot (\Phi_{\mathit{S}} \times d^{*\circ})) / c_2
\end{array}
\end{displaymath}
 \caption{Example model translated to FA}
  \label{fig:exfa}
\end{figure}

\section{Related work}
\label{sec:related}

Little work has been done concerning unbounded verification of Alloy
specifications. Arkoudas et al.~\cite{prioni} have developed a tool,
\texttt{Prioni}, which provides the axiomatization of the RL of Alloy
to Athena~\cite{athena}, a type-$\omega$ denotational proof
language. However, this translation is made to a first-order logic,
and thus the relational flavor of Alloy is lost. Frias et al. defined
a translation of Alloy formulas to FA~\cite{eqalloy}, from which the
initial inspiration for our translation was drawn. They then extended
the PVS~\cite{pvs} language to support FA, and used it to verify Alloy
specifications~\cite{friasall2}. Later, the same team extended the FA
so that it supports first-order quantifiers, making it more friendly
for Alloy users~\cite{dynamite}. They also presented a tool,
\texttt{DYNAMITE}, which provides interaction between PVS and Alloy
Analyzer. A significant methodological difference to our work is that,
in both these works, the focus was not on automatic verification but on
manual proofs assisted by a theorem prover.

Although initially inspired by~\cite{eqalloy}, our translation results
in much simpler expressions and ends up having very few similarities
with it. We also convert the input of the relations to a tuple
containing all quantified variables, from which the desired variable
is then selected. However, the way this goal is achieved differs
significantly from the original. In~\cite{eqalloy} all variable
occurrences were directly translated to a coreflexive which forced
the input to take the value of the desired variable. Like has been
presented, we expand the Alloy terms to RL, in order to be able to
completely remove the variables by applying some simple FA rules. Other
restrictions that highly increase the complexity of the initial
translation were lifted in our translation, like imposing every
formula to be of the shape $\top = R$ and enforcing that shape on the
inner terms. For comparison purposes, consider the following Alloy
formula:
\begin{center}
\verb+all a : A | some b : B | a in R.b && a in S.b+
\end{center}
Applying the translation defined in~\cite{eqalloy} results in the following formula:
\begin{align*}
&\overline{\overline{\overline{\overline{\top \cdot \rho (\langle id,
        \pi_2 \cdot \phi \rangle)} \cdot \langle \pi_1 \cdot \pi_1 ,
      \pi_2 \cdot \pi_1 , \pi_2 , \top \rangle \cdot \langle \pi_1 ,
      \pi_2 , \top \rangle } \cdot \langle id , \top \rangle} \cdot
  \top }  = \top 
\end{align*}
where
\begin{align*}
\phi = id \times \top \cap \overline{\delta(\pi_2 \cdot \pi_1 \cap \pi_2) \cap \overline{\rho(id \times R \cdot \delta(\pi_1 \cdot \pi_1 \cap \pi_2))} \cap id \times \top} \cap \\
\, \, \qquad \overline{\delta(\pi_2 \cdot \pi_1 \cap \pi_2) \cap
  \overline{\rho(id \times S \cdot \delta(\pi_1 \cdot \pi_1
    \cap \pi_2))} \cap id \times \top}
\end{align*}
and $\rho$ and $\delta$ compute the domain and range of a relation,
respectively (as a coreflexive).  On the other hand, our translation (without the
heuristic simplification) results in
\begin{equation*}
\top \subseteq \overline{\overline{(\top \cdot (\pi_1 \cap R \cdot \pi_2) \cap \top \cdot (\pi_1 \cap S \cdot \pi_2)) \cdot \langle id, \top \rangle} \cdot \top}
\end{equation*}
and with the heuristics just in $id \subseteq R^{\circ} \cdot S$.
In fact, the resulting expressions were so complex that the approach
proposed in~\cite{eqalloy} was abandoned by the authors
in~\cite{dynamite}. Another improvement from their work is that we
consider the whole Alloy model, including all type information from
declarations, while they consider only the translation of the
formulas. A feature of the approach presented in~\cite{friasall2} from which we
drew inspiration is the definition of the $n$-ary composition
operator.

Using Prover9 to verify relational models was already proposed in~\cite{crp9}. However, since they use only RA, their expressiveness power is limited to a 3 variable fragment of RL. A significant difference occurs in the way the types are represented. While they use \emph{vectors} to represent sets, and thus types, we use coreflexives. This change is motivated by our belief that coreflexives are more amenable to calculation. They also did not propose a translation from Alloy to RA and assume the model is already defined in this formalism. 

\section{Conclusions}
\label{sec:conc}

We have presented a framework for unbounded verification of Alloy
specifications with the automatic theorem prover Prover9. The key
ingredient of this framework is a translation from Alloy models into
FA, a point-free characterization of relational logic. This
translation greatly improves a previously existent one, by proposing a
different translation strategy and using heuristics to further
simplifies the results. It also covers the structural aspects of the
model, namely type declarations. These improvements made proof automation
with Prover9 viable.

The translation was fully implemented and tested on several examples
with complexity similar to our running example. On all these Prover9
was able to automatically verify the assertions. However, in order to
scale this approach to bigger case-studies some additional work is
still needed to find the appropriate set of axioms to be included in
Prover9. The main problem concerns the ubiquitous use in Alloy models
of relations with arity bigger than 2, and thus we are currently
searching for more useful laws characterizing the operators introduced
in Sect.~\ref{sec:nary}.

Although Prover9 was used to discharge the proofs, the translation presented is generic and not bound to it. For instance, since our translation provides much simpler results than the one from~\cite{eqalloy}, it could be interesting to use it as a replacement in order to increase the performance of the underlying proof system.

\bibliographystyle{splncs03}
\bibliography{allp9}

\appendix

\section{Grammar for a fragment of Alloy}
\label{sec:alloysyntax}

\begin{alltt}
\textit{Model}     \textit{:=} \textit{Paragraph*}
\textit{Paragraph} \textit{:=} \textit{Sig} \textit{|} \textit{Fact} \textit{|} \textit{Assert} 
\textit{Sig}       \textit{:=} \textit{[Mult]} \textit{[}abstract\textit{]} sig \textit{Id} \textit{[}extends \textit{Id}\textit{]} \{\textit{Rel*}\}
\textit{Fact}      \textit{:=} fact \{\textit{Form}\}
\textit{Assert}    \textit{:=} assert \{\textit{Form}\}
\textit{Rel}       \textit{:=} \textit{[Mult]} \textit{Id} \textit{|} \textit{[Mult]} \textit{Id} -> \textit{Rel}
\textit{Mult}      \textit{:=} some \textit{|} one \textit{|} lone \textit{|} set
\textit{Form}      \textit{:=} \textit{Exp} in \textit{Exp}  
           \textit{|} \textit{Form} && \textit{Form}  
           \textit{|} ! \textit{Form}
           \textit{|} all \textit{Id} : \textit{Exp} | \textit{Form}
           \textit{|} some \textit{Exp} | lone \textit{Exp}
\textit{Exp}       \textit{:=} \textit{Id}               -- relation identifier
           \textit{|} iden             -- identity relation
           \textit{|} none             -- empty relation
           \textit{|} univ             -- universe
           \textit{|} ~ \textit{Exp}            -- transposition
           \textit{|} \textit{Exp} . \textit{Exp}        -- composition
           \textit{|} \textit{Exp} & \textit{Exp}        -- intersection 
           \textit{|} \textit{Exp} + \textit{Exp}        -- union
           \textit{|} \textit{Exp} - \textit{Exp}        -- difference  
           \textit{|} \textit{Exp} -> \textit{Exp}       -- cartesian product
           \textit{|} \textit{Exp} <: \textit{Exp}       -- domain restriction 
           \textit{|} \textit{Exp} :> \textit{Exp}       -- range restriction
           \textit{|} * \textit{Exp}            -- reflexive-transitive closure
\end{alltt}

\end{document}